# Density dependence of the nuclear symmetry energy and neutron skin thickness in the KIDS framework[*]

P. Papakonstantinou[1]

[1] *Institute for Basic Science, Rare Isotope Science Project, Daejeon 34000, Republic of Korea*

**Abstract** The KIDS framework for the nuclear equation of state (EoS) and energy density functional (EDF) offers the possibility to explore systematically the effect of EoS parameters on predictions for a variety of observables. The EoS parameters can be varied independently of each other and independently of assumptions regarding the in-medium nucleon effective mass. Here I present a pilot study of the neutron skin thickness (NST) in nuclei of current interest. The results indicate that variations of the symmetry energy slope parameter L by roughly 10 MeV and variations of the droplet-model counterpart of the curvature parameter $K_\tau$ by roughly 20 MeV affect predictions by comparable amounts. However, structural details may also have sizable effects on predictions, notably in the cases of $^{68}$Ni and $^{208}$Pb. This work is part of a systematic investigation of the NST within the KIDS framework and of a broader effort to constrain the density dependence of the nuclear symmetry energy.

**Keywords** Symmetry energy, KIDS framework, neutron skin

## INTRODUCTION

The properties of very neutron rich nuclear systems are largely determined by the density dependence of the nuclear symmetry energy, $S(\rho)$ [1]. Recent and ongoing experiments aiming to measure the neutron skin thickness (NST) [2,3] and astronomical observations of neutron stars and gravitational waves [4] offer valuable information on the symmetry energy at sub-saturation and supra-saturation densities, respectively.

By convention, the density dependence of the symmetry energy is encoded in the values of its derivatives at saturation density, $\rho_0$. The following expressions for the energy per particle $\varepsilon(\rho,\delta)$ of zero-temperature unpolarized nuclear matter at density $\rho$ and isospin asymmetry $\delta$ summarize the necessary definitions:

$$\varepsilon(\rho,\delta) = E(\rho) + S(\rho)\delta^2 + \mathcal{O}(\delta^4),$$
$$E(\rho) = E_0 + K_0 x^2/2 + Q_0 x^3/6 + \mathcal{O}(x^4),$$
$$S(\rho) = J + Lx + K_{sym}x^2/2 + Q_{sym}x^3/6 + \mathcal{O}(x^4),$$

where $x=(\rho-\rho_0)/3$. Denoting the neutron and proton densities as $\rho_n$ and $\rho_p$, respectively, we have $\rho=\rho_n+\rho_p$ and $\delta=(\rho_n-\rho_p)/\rho$. There have been a great many studies of the lowest-order symmetry energy parameters J (value at saturation density) and L (slope parameter) taking advantage of data from a variety of observations, from basic nuclear structure and excitations to heavy ion collisions to compact stars. Based on the most recent work, the value of J lies most likely at 30-33 MeV and that of L between 40 and 65 MeV (but one may legitimately adopt values outside these intervals). The recently publicized PREX-II measurement of a rather thick neutron skin in $^{208}$Pb [3] has presented a puzzle as the result seems to point to a much higher value of L. That could throw off a host of predictions for neutron stars, the NST, and dipole polarizability values. On the other hand, the role of higher-order parameters of the symmetry energy, such as the curvature parameter $K_{sym}$, has not been much explored. In fact, it has been difficult to explore those parameters in an unbiased way, because in standard phenomenological approaches such as Skyrme and RMF models there are not enough free parameters to do so – see [5] for a discussion of this issue.

[*] Work supported by the Rare Isotope Science Project of the Institute for Basic Science funded by the Ministry of Science, ICT and Future Planning and the National Research Foundation (NRF) of Korea (2013M7A1A1075764).

The Korea-IBS-Daegu-SKKU (KIDS) theoretical framework for the nuclear equation of state (EoS) and energy density functional (EDF) [5-12] offers the possibility to explore all symmetry-energy parameters including higher-order ones independently of each other and independently of assumptions about the in-medium effective mass. Within KIDS, any set of EoS parameters can be transposed into an EDF in the highly convenient form of an extended Skyrme EDF and get tested in microscopic calculations of nuclear properties [9,10]. Related studies of symmetry-energy parameters based on astronomical observations and bulk nuclear properties were publicized recently [5,11]. As regards $K_{sym}$, the results in [5,11] point to a sizable correlation with neutron star radii. If we impose that all measurements of the radius of a canonical neutron star must be reproduced within their current uncertainties, we arrive at a likely value of $K_{sym}$ between roughly -150 and 0 MeV. The results in [11] point also to (a) a sizable correlation between the droplet-model counterpart of the curvature parameter, $K_\tau = K_{sym} - (6+Q_0/K_0)L$, with predictions for the NST in $^{208}$Pb and, quite significantly, (b) no correlation between predictions for the neutron star radius and the NST of $^{208}$Pb.

A dedicated KIDS study of the NST clearly is timely in light of the PREX-II measurement, other anticipated experimental data from, e.g., the CREX and R3B collaborations and the above findings. A Bayesian analysis of a variety of isovector nuclear properties within KIDS, including the NST and dipole polarizability, is in progress [12]. In the meantime, it is informative to explore further the correlation trends between symmetry energy parameters and predictions for the NST.

## KIDS EOS AND EDF PARAMETER SPACE

I consider a KIDS EoS with three independent parameters for symmetric nuclear matter and four independent parameters for neutron matter (equivalently, the symmetry energy), as has been found optimal [6,10]:

$$\varepsilon(\rho,0) = T(\rho,0) + c_0(0)\rho + c_1(0)\rho^{4/3} + c_2(0)\rho^{5/3},$$
$$\varepsilon(\rho,1) = T(\rho,1) + c_0(1)\rho + c_1(1)\rho^{4/3} + c_2(1)\rho^{5/3} + c_3(1)\rho^2,$$

where $T(\rho,\delta) \sim \rho^{2/3}$ denotes the kinetic energy per particle of a free Fermi gas. For more on the KIDS power expansion in terms of $\rho^{1/3}$ and its truncation see [6,10]. For symmetric nuclear matter I consider a standard EoS characterized by the saturation point $E_0 = -16$ MeV, $\rho_0 = 0.16$ fm$^{-3}$ and compression modulus $K_0 = 240$ MeV. These constants then determine the three coefficients $c_i(0)$. For the symmetry energy I consider all possible 7x8x9x6=3024 combinations of J, L, $K_{sym}$, $Q_{sym}$ values with

$$J = 30, 30.5, 31, \ldots, 33 \text{ MeV} ; L = 35, 40, 45, \ldots, 70 \text{ MeV} ;$$
$$K_{sym} = -160, -140, \ldots, 0 \text{ MeV} ; Q_{sym} = 0, 200, \ldots, 1000 \text{ MeV}.$$

The above then determine the remaining EoS expansion coefficients. The in-medium effective mass can also be freely varied without affecting the EoS parameters [5,9,12]. For the purposes of the present study, the isoscalar effective mass at saturation density is set to $\mu_s$=0.82 times the bare nucleon mass $m_N$, as generally favored by the energy of the giant quadrupole resonance in heavy nuclei. The isovector effective mass is set equal to $\mu_v$=0.82 times $m_N$ as well.

Next, the EoS including the effective mass values is transposed into an EDF of a generalized Skyrme form. Additional free paremeters exist which are not active in homogeneous matter, namely the isoscalar and isovector gradient coefficients $C_{12}$, $D_{12}$, and the spin-orbit coupling strength $W_0$ (see [10] for definitions). Here I assume the values $C_{12} = -67$ MeV fm$^5$ and $D_{12} = 10$ MeV fm$^5$. For $W_0$ I consider the five values

$$W_0 = 100, 110, \ldots, 140 \text{ MeV fm}^5.$$

So a total of 3024x5= 15120 EDFs are initially explored. The distribution of values for each parameter within the respective total interval is uniform in a coarse-grained partition, i.e., if we consider appropriately centered bins of width 0.5 MeV for J, 5 MeV for L, 20 MeV for $K_{sym}$, 200 MeV for $Q_{sym}$, and 10 MeV fm$^5$ for $W_0$. For the droplet parameter $K_\tau = K_{sym} - 4.446L$, the resulting distribution is shown on the left panel of Fig. 1.

Using a Skyrme-Hartree-Fock code, extended to accommodate the KIDS EDF, I calculate the properties of several closed-shell nuclei including their binding energies, charge root mean square radii $r_{ch}$, and the root-mean-square radii $r_p$ and $r_n$ of their point-proton and neutron density distributions. Next, I select a fraction of EDFs which reproduce well bulk nuclear properties. For this purpose I define the *average deviation per datum*,

$$\text{ADPD}(N) = \frac{1}{N}\sum_{i=1}^{N} \frac{|O_i^{calc} - O_i^{exp}|}{O_i^{exp}},$$

where O is an observable (here, binding energy or charge radius), the superscript "calc" denotes the calculated value and "exp" the value from an experimental measurement or evaluation, while N denotes the number of data considered. For each one of the 15120 EDFs defined above, I calculate a) the ADPD(6) corresponding to the energies and charge radii of the stable N=Z nuclei $^{16}$O, $^{40}$Ca and the energies of the unstable N=Z nuclei $^{56}$Ni, $^{100}$Sn (N=6 data in total) and b) the ADPD(19) corresponding to the energies and charge radii of $^{16}$O, $^{40,48}$Ca, $^{90}$Zr, $^{120,132}$Sn, $^{208}$Pb, and the energies of $^{56,68,78}$Ni, $^{100}$Sn, and $^{218}$U (N=19 data in total). Data for the binding energies and radii are taken from [13,14]. The values I find for ADPD(6) range from 0.7% to 1.6%, while values for ADPD(19) range from 0.41% to 2.4%. For the analysis that follows, I select the EDF sets for which ADPD(6)<1% and ADPD(19)<0.47%. Thus I select 855 sets of parameters as best performing in terms of bulk nuclear data based on ADPD(N).

Examining the distributions of J, L, $K_{sym}$, $Q_{sym}$ in the selected EoSs I find that they deviate only slightly from the initially pseudo-uniform distributions. Values of J between 31 and 33 MeV, L between 45 and 60 MeV, $K_{sym}$ from -150 and -20 MeV and $Q_{sym}$ > 500 MeV appear marginally more populated than other values based on the present sample. On the other hand, for $K_\tau$ the distribution of values within the selected EDFs, as shown on the right-hand side panel of Fig. 1, appears narrower than the initial distribution, with fringe values excluded and values in the neighborhood of roughly -300 to -370 MeV more populated.

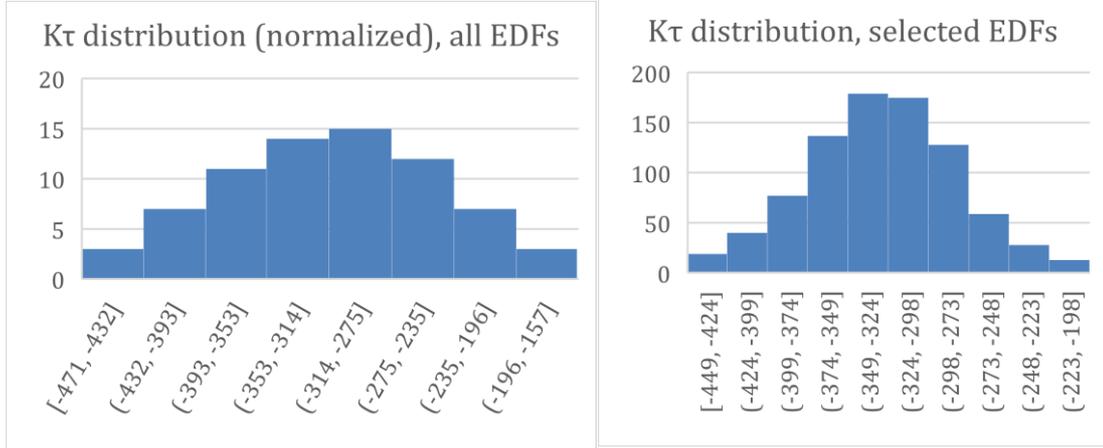

**Fig. 1:** The distribution of $K_\tau$ values before and after filtering the EDFs based on bulk nuclear data. The bin border values are given in units of MeV.

Finally I calculate the Pearson correlation coefficients $r_{XY}$, for the selected 855 parameter sets, where X and Y denote symmetry energy parameters:

$r_{JL}= 0.61$   $r_{JKsym} = 0.04$   $r_{JQsym} = 0.02$   $r_{LKsym} = 0.58$   $r_{LQsym} = -0.26$   $r_{KsymQsym} = 0.21$

Unsurprisingly, there are no prominent correlations with $K_{sym}$, $Q_{sym}$, since bulk nuclear properties probe a narrow regime of $S(\rho)$. For the lower-order parameters, linear fits yield

$J=0.059L + 28.6\text{MeV}$,  $L=-0.103K_{sym} + 61.6\text{MeV}$,  $L=-0.052K_\tau + 34.5\text{MeV}$,

which can be used as guidance for selecting reasonable parameter combinations but have no other significance. Different sets of data and selection criteria would give somewhat different relations as we will see also below.

## NEUTRON SKIN THICKNESS

I now present and discuss results for the NST, *i.e.*, the difference between the root-mean-square radii of the point-proton and point-neutron density distributions, $\Delta R_{np}$. I consider the stable nuclei $^{48}$Ca, $^{90}$Zr, $^{120}$Sn, and $^{208}$Pb and the β-unstable nuclei $^{68}$Ni, $^{78}$Ni, $^{132}$Sn. First, I calculate the correlation coefficients between the prediction for each nucleus' $\Delta R_{np}$ and each symmetry energy parameter. The results are shown in Table 1 and suggest that there is practically no correlation between $K_{sym}$ or $Q_{sym}$ and predictions for the neutron skin. On the other hand, $K_\tau$ may be at least as relevant as L. Higher J, higher L and lower $K_\tau$ all favor thicker neutron skins. I note that the correlation coefficients between the predictions for different nuclei were all found higher than 0.96.

|         | $^{48}$Ca | $^{90}$Zr | $^{120}$Sn | $^{208}$Pb | $^{68}$Ni | $^{78}$Ni | $^{132}$Sn |
|---------|-----------|-----------|------------|------------|-----------|-----------|------------|
| J       | 0.69      | 0.75      | 0.79       | 0.78       | 0.83      | 0.75      | 0.76       |
| L       | 0.50      | 0.53      | 0.59       | 0.62       | 0.66      | 0.62      | 0.64       |
| $K_{sym}$ | -0.13   | -0.10     | -0.10      | -0.11      | 0.01      | -0.02     | -0.07      |
| $Q_{sym}$ | 0.22    | 0.23      | 0.11       | 0.03       | 0.125     | 0.18      | 0.05       |
| $K_\tau$ | -0.62    | -0.61     | -0.68      | -0.71      | -0.61     | -0.60     | -0.69      |

**Table 1**: Correlation coefficients between predictions for $\Delta R_{np}$ of the indicated nuclei and symmetry energy parameters within a set of 855 EoS-EDFs as defined in the text.

In order to assess uncertainties in my eventual neutron-skin predictions, I perform the analysis with somewhat different filtering criteria for the EoSs. I assume the same EoS for symmetric nuclear matter as above, but

- fix the $Q_{sym}$ value to 583 MeV based on earlier fits to the Akmal-Pandharipande-Ravenhall EoS [6,10];
- for $\mu_s$, $\mu_v$ adopt respectively the values of 0.70 and 0.72, the latter value aiming at reproducing an isovector enhancement factor of 0.4; and
- vary J, L and for each EoS fit the gradient and spin-orbit parameters to 13 data, namely the energies and radii of $^{16}$O, $^{40,48}$Ca, $^{90}$Zr, $^{132}$Sn, $^{208}$Pb, and energy of $^{218}$U.

An inspection of results (which will not be presented here) for various J, L reveals a value of $K_\tau \approx$ -320 MeV as optimal for describing the above data, compatible with but not equal to the findings of the previous section. Considering the optimal combinations of J and L, one arrives at a relation of roughly the form J ≈ 0.05 L+29.8 MeV, which, for L≈30-70~MeV, gives for J estimates within approximately 1 MeV from what I obtained in the previous section. The optimal values for the gradient and spin-orbit terms are found to be $C_{12} \approx$ -75 MeV fm$^5$, $D_{12} \approx$ 30 MeV fm$^5$, $W_0$ = 120-130 MeV fm$^5$. $K_{sym}$ can be determined from L and $K_\tau$ in each case. Its values range from -186 to -9 MeV when L is varied from 30 to 70 MeV.

I will compare the results of both (J,L, $K_\tau$) analyses for the NST of nuclei of current interest, namely $^{48}$Ca, $^{68}$Ni, $^{132}$Sn, and $^{208}$Pb. Let me summarize the inputs considered next:

**Set A:** Fixed: $K_\tau$ = -320 MeV, $Q_{sym}$ = 583 MeV, $\mu_s$ = 0.70, $\mu_v$ = 0.72, $C_{12}$ = -75 MeV fm$^5$, $D_{12}$ = 30 MeV fm$^5$, $W_0$ = 130 MeV fm$^5$; varied: L = 30, 40,…, 70 MeV, J = 0.05 L + 29.8 MeV.

**Set B:** Same as Set A, except that $K_\tau$ = -340 MeV.

**Set C:** Fixed: $K_\tau$ = -340 MeV, $Q_{sym}$ = 600 MeV, $\mu_s$ = $\mu_v$ = 0.82, $C_{12}$ = -67 MeV fm$^5$, $D_{12}$ = 10 MeV fm$^5$, $W_0$ = 130 MeV fm$^5$; varied: L = 30, 40, …, 70 MeV, J = 0.059 L + 28.6 MeV.

**Set D:** Same as Set C, except that $Q_{sym}$ = 800 MeV.

The ADPD(19) value in all four cases is found lower than 0.5%.

The results for $\Delta R_{np}$ are shown in Table 2. Comparing sets A and B, one sees that a variation of $K_\tau$ by 20 MeV can account for about 0.01 fm variation in the prediction. The effect is tiny in terms of experimental precision, but comparable to the effect of varying L by 10 MeV.

Comparing sets C and D, one sees that a variation of $Q_{sym}$ by 200 MeV has roughly the same effect. Comparing sets B and C, one sees that different EDF assumptions, represented here by the effective mass and gradient terms, lead to a sizable effect especially in $^{68}$Ni and $^{208}$Pb. This result is in line with the observation in Ref. [9] regarding the influence of the effective mass value on predictions for $\Delta R_{np}$ of, among others, $^{48}$Ca, $^{132}$Sn, and $^{208}$Pb ($^{68}$Ni was not included): The predictions showed no obvious sensitivity except for $^{208}$Pb. An independent recent study also points to a dependence of the NST on structure details related to the effective mass [15].

|   | $^{48}$Ca | | | | $^{68}$Ni | | | | $^{132}$Sn | | | | $^{208}$Pb | | | |
|---|---|---|---|---|---|---|---|---|---|---|---|---|---|---|---|---|
| L | EoS/EDF Set | | | | EoS/EDF Set | | | | EoS/EDF Set | | | | EoS/EDF Set | | | |
|   | A | B | C | D | A | B | C | D | A | B | C | D | A | B | C | D |
| 30 | 0.14 | 0.15 | 0.16 | 0.16 | 0.09 | 0.09 | 0.14 | 0.14 | 0.19 | 0.20 | 0.21 | 0.21 | 0.08 | 0.09 | 0.14 | 0.15 |
| 40 | 0.15 | 0.16 | 0.16 | 0.17 | 0.10 | 0.11 | 0.15 | 0.15 | 0.20 | 0.21 | 0.22 | 0.23 | 0.09 | 0.10 | 0.15 | 0.16 |
| 50 | 0.16 | 0.16 | 0.17 | 0.17 | 0.11 | 0.12 | 0.16 | 0.16 | 0.21 | 0.22 | 0.23 | 0.24 | 0.11 | 0.12 | 0.16 | 0.17 |
| 60 | 0.16 | 0.17 | 0.17 | 0.18 | 0.13 | 0.13 | 0.16 | 0.17 | 0.23 | 0.24 | 0.25 | 0.26 | 0.12 | 0.13 | 0.18 | 0.19 |
| 70 | 0.17 | 0.18 | 0.18 | 0.19 | 0.14 | 0.15 | 0.17 | 0.18 | 0.24 | 0.25 | 0.26 | 0.27 | 0.13 | 0.15 | 0.19 | 0.20 |

**Table 2**. Predictions for the $\Delta R_{np}$ of $^{48}$Ca, $^{68}$Ni, $^{132}$Sn and $^{208}$Pb in units of fm for different values of the slope parameter L (shown in the leftmost column in units of MeV) and under different assumptions for the other symmetry-energy parameters and for constructing the EDF (see text).

**CONCLUSIONS**

A dedicated KIDS study of the neutron skin thickness is timely in light of the PREX-II measurement and other anticipated experimental data. I presented such a pilot study and predictions in nuclei of current interest. The results affirm the relevance of the slope and curvature parameters of the symmetry energy. However, structural details may also have sizable effects on predictions, notably in the cases of $^{68}$Ni and $^{208}$Pb.

**References**

[1] M. Baldo and G.F. Burgio, Prog. Part. Nucl. Phys. 91, p. 203 (2016); X. Roca-Maza and N. Paar, Prog. Part. Nucl. Phys. 101, p. 96 (2018).
[2] T. Aumann *et al.*, Phys. Rev. Lett. 119, 262501 (2017);
[3] D. Adhikari *et al.*, Phys. Rev. Lett. 126, 172502 (2021)
[4] B.P. Abbott *et al.*, Phys. Rev. X9, 011001 (2019); M.C. Miller *et al.*, arXiv: 2105.06979
[5] H. Gil, Y.-M. Kim, P. Papakonstantinou, and C. H. Hyun, Phys. Rev. C 103, 034330 (2021).
[6] P. Papakonstantinou, T.-S. Park, Y. Lim, and C. H. Hyun, Phys. Rev. C 97, 014312 (2018).
[7] G. Ahn and P. Papakonstantinou, HNPS Advances in Nuclear Physics vol.26, p. 112 (2019).
[8] P. Papakonstantinou and H. Gil, HNPS Advances in Nuclear Physics vol.26, p. 104 (2019).
[9] H. Gil, P. Papakonstantinou, C. H. Hyun, and Y. Oh, Phys. Rev. C 99, 064319 (2019).
[10] H. Gil, *et al.*, Phys. Rev. C 100, 014312 (2019).
[11] H. Gil, P. Papakonstantinou, and C. H. Hyun, arXiv: 2110.09802.
[12] Jun Xu and P. Papakonstantinou, in preparation.
[13] National Nuclear Data Center, Brookhaven National Laboratory, www.nndc.bnl.gov/nudat2/
[14] I. Angeli and K. P. Marinova, At. Data Nucl. Data Tables 99, 69 (2013).